\documentclass{sig-alternate-05-2015}
\usepackage{epsfig,endnotes,balance}
\usepackage{epstopdf}
\usepackage{enumerate}
\usepackage{amsmath}
\usepackage{wrapfig}

\usepackage{makecell}

\usepackage{algorithm,algorithmic}
\floatname{algorithm}{Protocol}
\usepackage{flexisym}
\usepackage{enumitem}
\usepackage{siunitx}

\usepackage{graphicx}
\usepackage{subcaption}

\usepackage{color}
\usepackage{multirow}
\graphicspath{../figures/}

\usepackage{etoolbox}
\makeatletter
\patchcmd{\maketitle}{\@copyrightspace}{}{}{}
\makeatother


\usepackage{tabularx}

\newcounter{protocol}

\author{
  Hongxiang Gu\\
  University of California, Los Angeles \\
  hxgu@cs.ucla.edu
  \and
  Miodrag Potkonjak\\
  University of California, Los Angeles \\
  miodrag@cs.ucla.edu
}

\title{Efficient and Secure Group Key Management in IoT using Multistage Interconnected PUF}

\begin{document}
\maketitle

\begin{abstract}
Secure group-oriented communication is crucial to a wide range of applications in Internet of Things (IoT). Security problems related to group-oriented communications in IoT-based applications placed in a privacy-sensitive environment have become a major concern along with the development of the technology. Unfortunately, many IoT devices are designed to be portable and light-weight; thus, their functionalities, including security modules, are heavily constrained by the limited energy resources (e.g., battery capacity). To address these problems, we propose a group key management scheme based on a novel physically unclonable function (PUF) design: multistage interconnected PUF (MIPUF) to secure group communications in an energy-constrained environment. Our design is capable of performing key management tasks such as key distribution, key storage and rekeying securely and efficiently. We show that our design is secure against multiple attack methods and our experimental results show that our design saves 47.33\% of energy globally comparing to state-of-the-art Elliptic-curve cryptography (ECC)-based key management scheme on average.
\end{abstract}

\section{Introduction}
Internet of Things (IoT) has been envisioned to be a revolutionary network that connects physical devices around us to perform intelligent tasks such as monitoring, communication, operation, and optimization. The advancement in IoT technology has enabled a wide spectrum of applications in a variety of environments to measure the various
environmental parameters \cite{guo2016enabling}. While IoT technology has greatly improved the efficiency and quality of our lives and works, various security challenges have become a major concern and doubt for further adoption of the technology. Security improvement in IoT system has become an increasingly popular topic in both academia and industry due to its urgency and profitability. In this paper, we are particularly interested in efficient and secure key management schemes in group communications in an IoT setting. 


Group communication through multicast/broadcast enables direct communication with the whole group, which is more efficient when compared to an equivalent unicast-based solution. Securing group communications consists of providing confidentiality, authenticity, and integrity of messages exchanged within the group \cite{veltri2013novel}. Among all security problems in IoT, group key management is one of the fundamentals in securing group communications. A group key essentially is a secret key shared by all members of a group so that all group communication packages are encrypted before they are being transmitted using this group key. An unauthorized user may receive group communication packages due to network error or intentional interception, however, without the right group key, the illegal user cannot decrypt the received packages. 

Group key management schemes in IP networks, though have been studied for decades, cannot be directly applied to IoT as IoT devices are heavily constrained by the limited resource and energy capacity. Limited resources impose new challenges regarding storage and computation requirements, meaning each node is incapable of storing a large key database or conduct heavy cryptographic computation. The energy constraint additionally requires key verification and computation procedures to be energy efficient. 

For the above two reasons, physically unclonable functions (PUFs), a type of low-power security primitive with unclonable and unpredictable properties, naturally appears as an ideal solution to the problem. In this paper, we propose to apply a novel low power PUF structure called Multistage Interconnected PUF (MIPUF) to the domain of group key management in IoT. We believe the low power and unclonable, unpredictable nature of MIPUF not only improves the security of group key management protocols but also meet the tighter energy requirements on IoT nodes. Our design of interconnection reconfiguration in MIPUF is robust and secure against modeling attacks by changing the challenge-response mapping. The group key is stored and managed by a new set of PUF functions every time we reconfigure the MIPUF in every IoT device, creating an additional layer of security and protection. We also show that our key management scheme including key distribution, key storage and rekeying is resilient against a wide range of attacks.  Lastly, we show that our group key management protocol is power and energy efficient. Our simulation results show that we are 47.33\% more energy efficient comparing to state-of-the-art ECC-based key management schemes.

\section{Related Work}
Several efforts have been made in creating efficient group key management protocols for group communications in IoT and wireless sensor networks (WSN) to meet the energy and computation constraints. Notably, Zhu et al. proposed an efficient security mechanism for large-scale distributed sensor networks \cite{zhu2006leap+}. Roman et al. analyzed the applicability of public-key cryptography based protocols and link-layer oriented key management systems (KMS) in IoT settings \cite{roman2011key}. Abdallah et al. proposed a novel efficient and scalable key management mechanism for wireless sensor networks and proposed to reduce power and energy consumption by using ECC \cite{abdallah2015efficient}. All work listed above utilizes expensive cryptographic primitives to secure their group key management protocols without investigating the possibility of utilizing some novel low-power hardware security primitives to meet the energy requirements.

Recently, PUFs, as a popular type of low-power security primitive, have been proposed to be used in a number of key management subtasks in IoT settings. Gu et al. proposed multiple PUF architectures that provides compatibility to secure authentication. \cite{gu2018securing} \cite{xu2016ultra} \cite{gu2016energy}. Mukhopadhyay proposed a novel device authentication method that takes the advantage of the unclonable property of PUFs \cite{mukhopadhyay2016pufs}.  Most recently Huang et al. investigated a key distribution protocol assisted using ring oscillator PUFs (ROPUFs)\cite{huang2017puf} which significantly reduces the storage overhead and latency for securely distributing secret keys. Unfortunately, these works only focus on a specific subtask of key management and fail to provide detailed security or overhead analysis. We differentiate ourselves by design a novel PUF architecture that can be applied to the entire key management lifecycle including key distribution, key storage and rekeying in IoT. We have also performed a security and overhead analysis to prove that our work is both secure and efficient. 

\section{Multistage Interconnected PUF} \label{mipuf}
In this section, we propose a novel PUF structure called Multistage Interconnected PUF (MIPUF). We borrow the idea of multistage interconnection networks (MINs) from computer networks field. MINs allow the processing elements (PEs) to be interconnected using Switching Elements (SEs) such that the interconnection provides high configurability and speed with low cost. We propose to use such structure to interconnect PUFs so that the interconnected PUFs can be configured easily. The interconnected PUFs significantly increase the system complexity as well as break the linearity, resulting in increased difficulty in modeling the system. The configurability also allows the challenge-response pairs (CRPs) of the network to be remapped from time to time, protecting the system from modeling attacks. 

\subsection{Processing Elements (PEs)}
We name the PE in a MIPUF a MIPUF node. A MUPUF node is the most fundamental building block of the network. A MIPUF node is a single or a group of strong PUFs that take an $n$-bit challenge and generate an $m$-bit response. A strong PUF is defined as a PUF that supports a large number of CRPs and there exists a number of implementation. For the sake of implementation easiness, our implementation of a MIPUF node consists of $m$ $n$-bit arbiter PUFs running in parallel and sharing the same pulse signal and challenge vector. Even though arbiter PUFs are known to be weak against various modeling attack, our experimental results show that multistage interconnection significantly improves the resilience against them. In this paper, we merely use MIPUF node implemented with arbiter PUFs as an illustrative example and a proof of concept. Security properties of MIPUF implemented using more advanced strong PUFs such as LRR-DPUF  \cite{miao2016lrr} are expected to exceed our collected results. 

\subsection{Switching Elements (SEs)}
Similar to the concept of SEs in computer networks, the SE in MIPUF serves as a way to route and switch signals. In our implementation, an SE is a set of multiplexers that switch or not switch $n$ signals based on a configuration bit. In our case, we use SEs to connect the response of a previous MIPUF node to the next node as the new challenge. The SEs between two nodes are controlled by a configuration vector.

\subsection{Multistage Interconnection}
\begin{figure}[h!]
    \centering
    \includegraphics[width=2in]{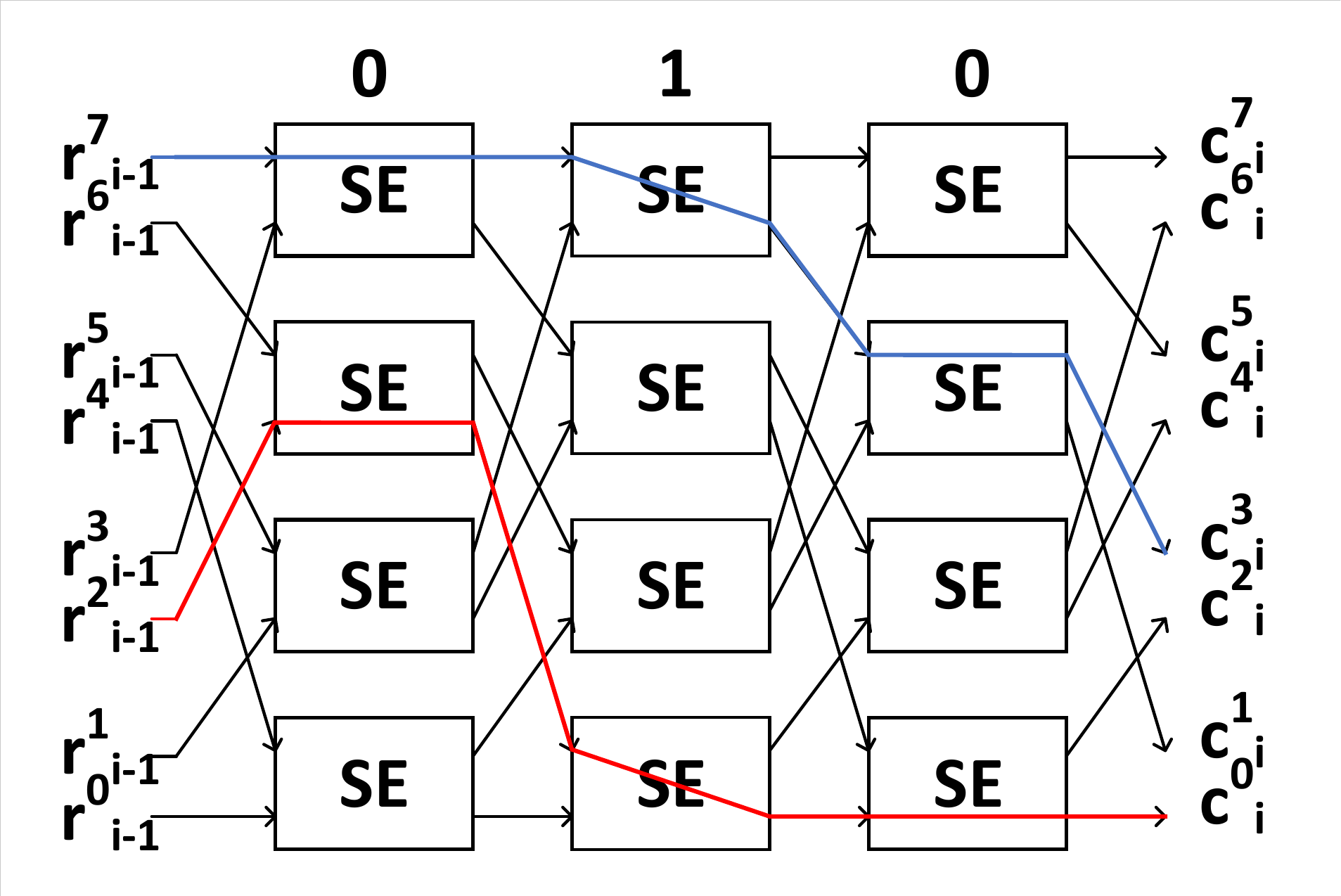}
     \setlength{\belowcaptionskip}{-12pt}
    \caption{Network structure in Omega network style. $r^{k}_{i-1}$ indicates the $k$-th response (output) bit of the $(i-1)$-th MIPUF node, $c^{k}_{i}$ indicates the $k$-th challenge (input) bit of the $i$-th MIPUF node.}
    \label{fig:network_struct}
\end{figure}

Multistage interconnection networks find a balance between the cost and configurability. We believe a blocking multistage connection is the most cost-efficient for MIPUF implementation and provides sufficient configurability. A blocking multistage connection cannot realize all possible connections between inputs and outputs since a connection between one free input to another free output is blocked by an existing connection in a network; however, it is much cheaper to implement. We propose to implement a blocking interconnection in a MIPUF in an Omega network style \cite{lawrie1975access} which consists of 2$\times$2 SEs. Each input has a dedicated connection to an output, providing $2^N$ different switchings and having a complexity of O(N log(N)) for an N$\times$N connection between two MIPUF nodes. An example of such interconnection is shown in Figure \ref{fig:network_struct}. To be noted that we do not allow port rearrangement in MIPUF, each input should be routed to a unique output and each output should be directed from a unique input given a specific configuration. 

\subsection{Protecting Network Configuration}
The signal routing between any two MIPUF nodes is controlled by a configuration vector. We propose to secure the configuration vectors using existing MIPUF nodes in the system so that the real interconnection configuration remains hidden. The configuration vector for the interconnection between node $i$  and $i+1$ depends on the encrypted result of the user provided configuration bits using the nodes from node 1 to node $i-1$, for $i>1$. To note that we use MIPUF nodes in the previous levels to encrypt SE configurations to reduce correlation between the output of a node and it's immediate SEs. The user provided configuration is passed to the SEs between the first two nodes and propagate along the network to configure the remaining SEs connected to the later nodes.  An attacker or even the user who provided the configuration, cannot obtain information on the real interconnection between MIPUF nodes without characterizing each node. Besides, we also significantly reduce the number of bits an user needs to provide. In a key management protocol, our proposed method could also significantly reduce the communication cost. 

\subsection{Security Evaluation of MIPUF}
\subsubsection{Uniqueness and Reliability}
Two most important properties of PUFs are uniqueness and reliability. Uniqueness means that the responses for a specific PUF design implemented on different devices should be significantly different when provided with the same challenge. Reliability indicates the response should be stable enough when repeating the same challenge on the same device. Since our design of MIPUF depends on existing PUF implementations, our focus is that our multistage interconnection does not compromise the security properties of the PUF implementation we depend on. We modify the definition of uniqueness and reliability as follows.

\begin{itemize}
\item \textit{Inter-configuration variation (uniqueness)}. How many MIPUF output bits are different between two different configurations of the same MIPUF? Ideally, this variation should be 50\%.
\item \textit{Intra-configuration variation (reliability)}. How many MIPUF output bits differs when re-generated again from a MIPUF with a specific configuration? Ideally, this variation should be 0\%.
\end{itemize}

We directly compare these two metrics with intra-chip variation and inter-chip variation metrics in regular PUF evaluations.  As a proof of concept, we compare our arbiter PUF based MIPUF implemented using arbiter PUFs with regular FPGA-based arbiter PUFs implemented on five different FPGAs. The results are collected from Xilinx Spartan-6 XC6SLX45 platform using the implementation described in \cite{zheng2014secure}. 

\begin{figure}[h!]
    \centering
        \begin{subfigure}[b]{0.4\textwidth}
        \includegraphics[width=\textwidth]{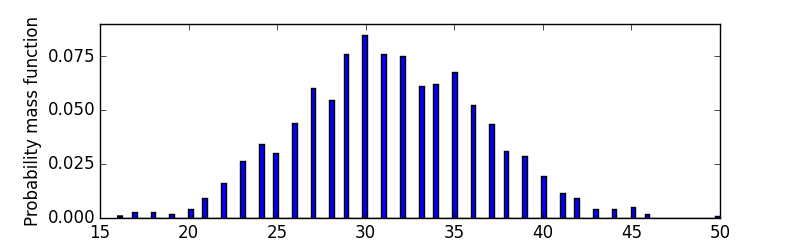}
        \caption{Inter-configuration variation for MIPUF is 47.9\% (Avg = 30.7 bits / 64 bits).}
        \label{fig:inter}
    \end{subfigure}
     \begin{subfigure}[b]{0.4\textwidth}
        \includegraphics[width=\textwidth]{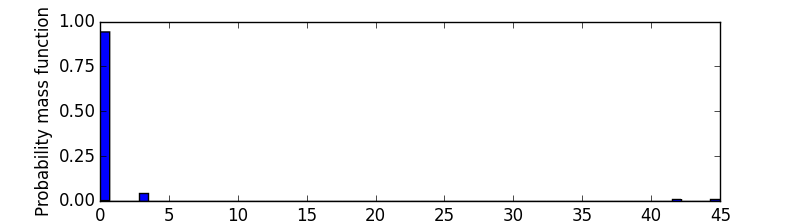}
        \caption{Intra-configuration variation for MIPUF with fuzzy extractor is  2.67\% (Avg = 1.71 bits / 64 bits). Environment range from \SI{20}{\celsius}, 0.95V to \SI{65}{\celsius}, 1.2V.}
        \label{fig:intra}
    \end{subfigure}
            \setlength{\belowcaptionskip}{-10pt}
         \caption{Inter-configuration and intra-configuration variation of a MIPUF with four nodes. Each each node is implemented using 64 32-bit arbiter PUFs. The interconnection between nodes is designed in a blocking fashion as shown in Figure \ref{fig:network_struct}.}
         \vspace*{-2mm}
\end{figure}

Figure \ref{fig:inter} illustrate the probability distribution of the inter-confi -guration variation of a MIPUF. The x-axis is the number of output bits that are different between two different interconnection configurations; the y-axis is the probability. The bars show experimental results collected on 1,225 pairs of outputs collected from 50 different configurations. Our experiment results (47.9\%) is very close to the ideal case of 50\%. Our results even show a slight improvement comparing to the inter-chip variation of arbiter PUFs implemented on FPGAs (47.0\%). 

We calculated a 35.37\% intra-configuration variation when no error correction is applied. Consider the intra-chip variation of 64 128-bit arbiter PUFs implemented on five different FPGAs is only as little as 2.90\%, MIPUF is very unstable without error correction. The reason is simple and intuitive, as all MIPUF nodes are connected in such a way that each node takes the output of the previous node as the input, an error in the first node could result in avalanche effect in intra-configuration variation. Thus, we propose to use a lightweight fuzzy extractor between every MIPUF node as an error correction mechanism \cite{herder2017trapdoor}, and the resulting intra-variation is significantly reduced to 2.67\%. Since a MIPUF with $n$ nodes requires $n$ clock cycles to generate the result, the fuzzy extractors can be shared for all node outputs. 

Figure \ref{fig:intra} illustrates the probability distribution of the intra-configuration variation of the same MIPUF. The environments parameters ranging from \SI{20}{\celsius}, 0.95V to \SI{65}{\celsius}, 1.2V. The bars show experimental results collected on 50,000 different random interconnections. Each configuration is performed on 10,000 challenges and repeated 20 times. Noted that the major contributors to the intra-configuration variation are two extremely rare ($<$ 0.5\%) case of Hamming distance greater than 40. 

\subsubsection{Resilience Against Modeling Attack} \label{resilience}
Several PUF-based systems are vulnerable to a variety of modeling attacks \cite{ruhrmair2013puf}. We observed that MIPUF significantly boost modeling attack resilience by increasing the system complexity and breaking the system linearity. Table \ref{table:realresult} shows the best prediction accuracy on MIPUF vs. a variety of PUFs implemented on FPGA using attack approaches described in \cite{ruhrmair2013puf}.  We observe that all prediction accuracies for a single-bit in MIPUF outperforms other designs and are all close to the ideal case of 50\%.

\begin{table}[ht]
    \centering
    \begin{tabular}{| l | l | l | l |}
        \hline
        Architecture & LR & ES & DL\\ \hline
        \textbf{MIPUF - 4 nodes} & \textbf{51.33\%} & \textbf{59.18\%}   & \textbf{50.59\%} \\\hline
        256-bit 4-XOR PUF  & 97.21\% & 76.02\%   & 78.42\%\\\hline
        1024-bit arbiter PUF & 96.57\% & 98.28\%   & 88.98\% \\\hline
        1024-bit 64-ff PUF    & 58.29\%  & 95.68\%   & 87.70\% \\\hline
    \end{tabular}
    \caption{Best single-bit prediction accuracy on different PUF architectures using logistic regression (LR), evolution strategies (ES) and deep learning (DL) attacks out of 100 runs. Each attack uses 100,000 CRPs. Total number of arbiter PUF segments used in all architectures are fixed to 1,024.}
    \label{table:realresult}
\end{table}

In addition to high resilience against modeling attacks, MIPUF also allows cheap and fast reconfiguration. Frequent reconfiguration of MIPUF renders modeling attacks almost impossible. We investigate this topic in more detail in Section \ref{resilience}.

\section{Group Key Management}
In this section we show that we can utilize the MIPUF structure to securely establish a group key management protocol with three major components, respectively key distribution, key storage and rekeying. Key distribution is the process to securely deliver the shared secret key to every authorized group member. After the group key has been successfully distributed, the most important task would be to securely store the secret key so that the user could easily access the key when needed, but an adversarial is forbid to peek or tamper with the secret key. Lastly, rekeying allows a group to renew or replace the group key from time to time. 

To illustrate our protocol, we first define an IoT model consists of a control unit with higher computational power and multiple IoT device/nodes that are constraint by both computational power and battery life. Each IoT node embeds a MIPUF, a hardware hashing function and a very compact AES implementation. 

\subsection{Key Distribution}\label{keydistribute}
According to the model described above, a well-designed group key distribution protocol is proposed. The protocol is shown in Protocol \ref{keydisprotocol}. For each node, a group key can be delivered securely with exchange of two messages.

\begin{algorithm}[!ht]
\begin{itemize}
\item Input: A list of group member in group $\mathcal{G} = \{N_0 \cdot \cdot \cdot N_n\} \subseteq \mathcal{N}$ $n$ being the total number of IoT nodes in the group. A random group key $key_g$.
\item Goal: Securely deliver $key_g$ to all $N_i \in \mathcal{G}$.
\end{itemize}
\begin{enumerate} [topsep=0pt,itemsep=-1ex,partopsep=1ex,parsep=1ex]
  \item \textbf{Preliminary Phase}
  \begin{enumerate}[topsep=0pt,itemsep=-1ex,partopsep=1ex,parsep=1ex]
    \item Before the deployment of an IoT node, the control unit assigns a unique ID ($N_i$) to it. Initially the node derives the interconnection configuration $\gamma_i = H(N_i)$ from $N_i$ and generates a CRP ($c^{\gamma_i}_i$, $r^{\gamma_i}_i$) using the MIPUF $\mathbb{F}_i$ where $r^{\gamma_i}_i = \mathbb{F}^{\gamma_i}_i(c^{\gamma_i}_i)$.
    \item The control unit securely store the tuple ($\gamma_i$, $c^{\gamma_i}_i$, $r^{\gamma_i}_i$) in the database, and node $N_i$ securely stores $c^{\gamma_i}_i$ and $\gamma_i$.
  \end{enumerate}

  \item \textbf{Key Delivery Phase}
  \begin{enumerate}[topsep=0pt,itemsep=-1ex,partopsep=1ex,parsep=1ex]
    \item When a group  $\mathcal{G}$ is formed, the control unit first check if all group members exists based on the unique ID. If not, the protocol is aborted.
    \item For IoT node $N_i \in \mathcal{G}$, the control unit generates a random new configuration $\gamma^{\prime}_i$.
    \item For IoT node $N_i \in \mathcal{G}$, a group key hint $p_i = r^{\gamma_i}_i \otimes key_g$ and a new configuration hint $f_i = r^{\gamma_i}_i \otimes \gamma^{\prime}_i$ are generated.
    \item An encrypted message $msg^k_i$ containing $p_i$ and $f_i$ is transmitted using unicast to each group member $N_i \in \mathcal{G}$. $msg^k_i = \{E_{r^{\gamma_i}_i}(N_i \| p_i  \| f_i) \| H(N_i \| c^{\gamma_i}_i)\}$, ``$\|$" indicates the concatenation operation, E is the encryption operation using AES and H is a hashing operation. 
    \end{enumerate}

  \item \textbf{CRP update Phase}
  \begin{enumerate}[topsep=0pt,itemsep=-1ex,partopsep=1ex,parsep=1ex]
  \item Upon receiving $msg^k_i$, IoT node $N_i$ first decrypts the message using $r^{\gamma_i}_i$: $D_{r^{\gamma_i}_i}(E_{r^{\gamma_i}_i}(N_i \| p_i  \| f_i))$, D being the AES decryption operation. $N_i$ verifies the validity of the message by comparing the hash $H(N_i \| c^{\gamma_i}_i)$. If there exists a mismatch in the hash, report error to the control unit.
  \item The group key $key_g$ and the new interconnection $\gamma^{\prime}_i$ are derived from $p_i$ and $f_i$ decrypted from the decrypted $msg^k_i$.
  \item $N_i$ generates a new CRP ($c^{\gamma^{\prime}_i}_i$, $r^{\gamma^{\prime}_i}_i$) where $c^{\gamma^{\prime}_i}_i = H(c^{\gamma_i}_i)$, $r^{\gamma^{\prime}_i}_i = \mathbb{F}^{\gamma^{\prime}_i}_i(c^{\gamma^{\prime}_i}_i)$. $N_i$ sends an encrypted message $msg^u_i = E_{r^{\gamma_i}_i}(N_i \| c^{\gamma^{\prime}_i}_i \| r^{\gamma^{\prime}_i}_i)$ back to the control unit. 
  \item The control unit decrypt $msg^u_i$ using $r^{\gamma}_i$ and updates the database by replacing the tuple ($\gamma_i$, $c^{\gamma_i}_i$, $r^{\gamma_i}_i) with (\gamma^{\prime}_i$, $c^{\gamma^{\prime}_i}_i$, $r^{\gamma^{\prime}_i}_i$). If the control unit has not received an update message $msg^u_i$ after some predefined timeout or $c^{\gamma^{\prime}_i}_i = H(c^{\gamma_i}_i)$, an abort is called.
      \end{enumerate}
\end{enumerate}
\caption{Group Key Distribution Protocol}\label{keydisprotocol}
\end{algorithm}

\subsection{Key Storage}
After the key distribution, the group key can be extracted from the MIPUF when the correct challenge and interconnection configuration are provided. Unlike other crypto-based key management systems, we do not directly store the group key in the memory. Instead, the group key is extracted on the fly from the group key hint $p_i$. We believe this approach is secure because an attacker can only obtain the real group key if he has access to both the storage (containing $p_i$, $f_i$ and $c^{\gamma_i}_i$) and the MIPUF ($\mathbb{F}^{\gamma_i}$, compromising either the storage or the MIPUF does not compromise the security of the whole design. Also, the group key is only used upon receiving or transmitting group messages, thus storing the real key using low-power MIPUF is also highly energy efficient.

\subsection{Rekeying}
Group keys need to be regenerated, redistributed or updated whenever there is a dynamic change to the group to preserve security. One important motivation to rekey is that groups are not always static. When a member leaves the group, it should not be able to decrypt future group communications (\textbf{forward security}); when a new member joins, it should not be able to decrypt past group communications (\textbf{backward security}). Group key should also be completely rekeyed when potential leakage is detected for security considerations. Here we discuss all three possible cases. 

\subsubsection{New Member Joins the Group}
Without loss of generality, we assume a new IoT node $N_{\alpha}$ intend to join a group $\mathcal{G}$, $N_{\alpha} \notin \mathcal{G}$. For efficiency considerations, redistributing a new key to all group members is expensive and inefficient. Instead, we propose to use the current secret to encrypt the new group key and this process is leakage free. Specifically, the control unit sends out a message $msg^{join} = \{E_{key_g}(key^{\prime}_g)\}$ to $\forall N_i \in \mathcal{G}$. The existing group members calculate and store the new group key hint $p^{\prime}_i = r^{\gamma_i}_i \otimes key^{\prime}_g$ and deletes $key^{\prime}_g$ upon receiving and decrypting $msg^{join}_i$. The new member will have to complete the whole key distribution process described in Section \ref{keydistribute}. Backward security is preserved using this method since the new member has no information about the old group key.

\subsubsection{Existing Member Leaves the Group}
Removing an existing member from the group is more complicated than adding a new member. Here we propose to divide group $\mathcal{G}$ into $m$ subgroups $\mathrm{g}_j \subset \mathcal{G} = \{\mathrm{g}_1 \cdot \cdot \cdot \mathrm{g}_m\}, 1\leq j \leq m$. All nodes in the same subgroup share the same interconnection configuration $\gamma_{\mathrm{g_j}}$. Again, without losing the generality, we assume an IoT node $N_{\beta} \in \mathrm{g}_j \subset \mathcal{G}$ intend to leave the subgroup where all members in the subgroup use the same MIPUF interconnection configuration $\gamma_j$. The control node first multicast/broadcast $m-1$ messages $msg^{leave}_i= \{E_{\gamma_i}(key^{\prime}_g \| H(\gamma_i)\}$ containing the new key to all the subgroups encrypted using the configuration $\gamma_i, i\neq j, 1 \leq i \leq m$. Upon receiving the message, each node first decrypts the message using its own configuration $\gamma_i$ and check if H($\gamma_i$) matches the one in the decrypted $msg^{leave}_i$. If so then the decrypted new group key $key^{\prime}_g$ is valid, otherwise, discard the message. No member of $\mathrm{g}_j$ including the leaving node have any knowledge of the configurations of other subgroups, thus incapable of decrypting the message correctly. The control unit should then perform unicast communications to all members of $\mathrm{g}_j$ by distributing the new group key $key^{\prime}_g$ and a new configuration $\gamma^{\prime}_j$ to replace $\gamma_j$.

\subsubsection{Complete Rekeying} \label{resilience}
MIPUF can still be modeled if a significantly large enough set of CRPs is collected. However, MIPUF can be reconfigured to neutralize modeling attacks by completely remap the input-output mapping. We propose to perform a full rekeying once the total number of CRPs generated exceeds a calculated sample complexity lower bound that equals to the sufficient training set size to break the MIPUF. Equation \ref{eq:reipn} describe a sample size lower bound in terms of the IPN model parameters, where $m$ is the number of nodes in MIPUF and $n$ is the maximum number of PUFs in a MIPUF node. $k = VC(\mathbb{F})$ where VC is the Vapnik-Chervonenkis-dimension and $\mathbb{F}$ is largest single PUF in MIPUF. $\delta$ is the failure probability and $\epsilon$ is the learning error.
\begin{equation}
\text{Sample complexity} \sim \frac{(m \cdot k + m) \cdot n + ln(\frac{1}{\delta})}{\epsilon}
\label{eq:reipn}
\end{equation}

\section{Evaluation}
\subsection{Security Analysis}
We make the following assumptions for our security analysis. The physical security of the MIPUF is secured; however, an attacker is allowed query the CRPs as much as needed. The wireless channels used for communication are not secured after the initial preliminary phase. The hash function and compact AES on each node are secure. The control unit key database is secure. We summarize our security analysis against several popular attacks as below:

\textbf{Eavesdropping Attack}: During the key distributing and rekeying process, all messages containing $\gamma_i$, $c^{\gamma_i}_i$, $r^{\gamma_i}_i$ or $key_g$ are encrypted by AES. Thus eavesdropping attack is invalid.

\textbf{Man-in-the-middle Attack}: Before updating the new CRP in $\textbf{3c}$ in Protocol \ref{keydisprotocol}, the new challenge is a one-way hash of the previous challenge which is checked by the control unit, thus render the attack useless.

\textbf{Replay Attack}: Neither the IoT node nor the control unit would be able to correctly decrypt a message encrypted using a previous response since old responses are discarded after the update.Thus the hash check in $\textbf{3a}$,  $\textbf{3d}$ in Protocol \ref{keydisprotocol} would fail during key distribution. Forward security in the rekeying process is designed to protect the system from such attacks.

\textbf{Impersonation  Attack}: Based on our assumption, the preliminary phase is secure thus the initial key and MIPUF configuration are secured. Also, the modeling attack resilience and the reconfigurability of MIPUF prevents an attacker to impersonate an IoT node even if we allow him to query the CRPs.

Comparing to other designs (e.g. ECC-based scheme) our group key management scheme enjoys at least the same level of security while adding an additional level of security at physical level utilizing the unclonability properties of MIPUF.

\subsection{Overhead Evaluation} \label{overhead}
In this section we assume that there are $N$ nodes in the group and the group is split into $M$ subgroups. The MIPUF we use in each IoT nodes takes an $\textbf{a}$-bit challenge, a $\textbf{b}$-bit configuration vector and generates a $\textbf{c}$-bit output (assuming $\textbf{b} \geq \textbf{c}$ ). Node ID is a $\textbf{l}$-bit vector. The hash function hashes any input to an $\textbf{a}$-bit string. The group key has length of $\textbf{c}$-bits. The random number generator cost $E_R$ units of energy per operation. The energy consumption of MIPUF, hash function, random number generator, XOR operation and the very compact AES are \boldmath{$E_{P}$}, \boldmath{$E_H$}, \boldmath{$E_R$}, \boldmath{$E_X$} and \boldmath{$E_A$}. 

$\textbf{Communication Cost}$: The length of messages: $msg^k_i$, $msg^u_i$, $msg^{join}$ and $msg^{leave}_i$ are: $\textbf{a}+\textbf{b}+\textbf{c}+\textbf{l}$, $\textbf{a}+\textbf{c}+\textbf{l}$, $\textbf{a+c}$ and $\textbf{a+c}$. Thus the total number of messages need to be sent for key distribution and node join/leave rekeying are: $\textbf{N}$,  $\textbf{3}$ and $(\frac{\textbf{2N}}{\textbf{M}}\textbf{-2})+(\textbf{M-1})$. For node leave rekeying, minimum cost is achieved when $M = \sqrt{N}$.

$\textbf{Storage Overhead}$: The control unit stores the Node ID, CRPs and the current configuration of all nodes; thus the storage overhead at the control unit is $N\cdot$$\textbf{(a+b+2c+l)}$ bits. Each IoT node stores the Node ID, current challenge, current configuration and the group key hint which has a storage overhead of $\textbf{a+b+c+l}$ bits.

$\textbf{Energy Cost}$: The control unit spends $2E_A$+ $2E_H$ + $2E_R$+ $2E_X$ units of energy to distribute the group key to one node. Each node spends $2E_A + 2E_H + E_P + 2E_R$ to receive the key and update the CRP. During member join rekeying, the control units spends $E_A+E_R$ units of energy to update the group key to existing members and $2E_A + 2E_H + 2E_R + 2E_X$ to the new member. The new node spends $2E_A + 2E_H + E_P + 2E_X$ and old members spend $E_A+E_X$ units of energy respectively. During member leave rekeying, the control unit spends $(M-1)\cdot(E_A+E_H+E_R)$ units of energy to update the group key to existing members that are not in the same subgroup as the leaving node and $(\frac{N}{M}-1)\cdot(2E_A + 2E_H + 2E_R + 2E_X)$ to the update all members in the same subgroup. All members that are in and not in the same subgroup as the leaving node spends $(M-1)\cdot(E_A+E_H+E_R)$ and $E_A+E_H$  units of energy.

\begin{figure}[h!]
    \centering
    \includegraphics[width=2.3in]{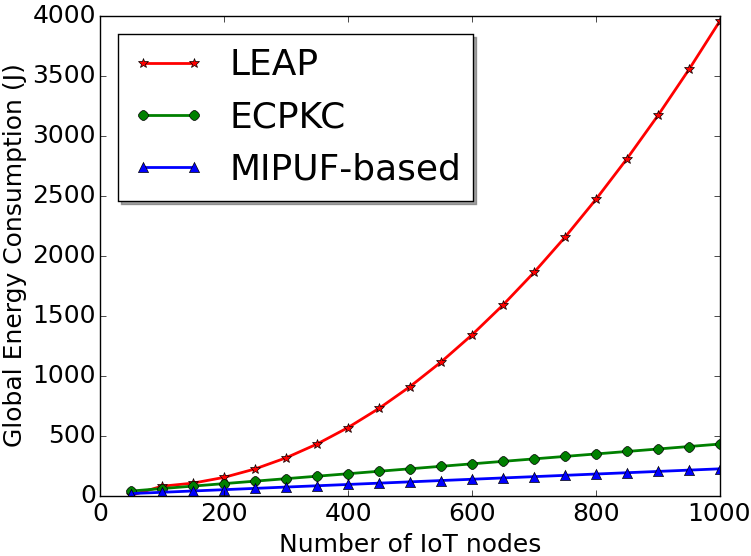}
    \caption{Simulated global energy consumption (J) vs. total number of IoT nodes under the settings introduced in \cite{abdallah2015efficient}.}
    \label{fig:global}
            \vspace*{-6mm}
\end{figure}

We compare our global energy consumption to two other key management protocols: Localized Encryption and authentication protocol (LEAP) \cite{zhu2006leap+} and Elliptic Curve Public Key Cryptography (ECPKC) \cite{abdallah2015efficient} in simulation using the parameters described in \cite{abdallah2015efficient} for a fair comparison. The energy consumption parameters for our design are estimated from our implementation described in Section \ref{impres}. The comparison for simulated results for global energy consumption for three key management schemes can be seen in Figure \ref{fig:global}. LEAP uses significantly much more energy than both ECPKC and our proposed scheme as it grows quadratically. The global energy consumption of both ECPKC and our proposed scheme grows linearly. Since our proposed design uses low-power MIPUF instead of energy-hungry ECC to achieve power efficiency. We observe that our proposed scheme uses about 47.33\% less energy for key distribution. 

\subsection{Implementation Results}\label{impres}
We implemented our key management hardware support for IoT nodes on Xilinx Spartan-6 LX45 FPGAs to measure the area and power. Our implementation consists of a MIPUF with three nodes each consists of 128 32-bit Arbiter PUFs. The fuzzy extractor, the hash function and the AES module are implemented based on \cite{herder2017trapdoor} \cite{sha1fpga} and \cite{chodowiec2003very} accordingly. Table \ref{table:resource} shows the area and power overhead break down of our implementation. Our MIPUF seems to be more expensive due to FPGA-based arbiter PUF implementations are known to be inefficient. The overhead of MIPUF and the hardware support are expected to be significantly reduced if implemented on ASIC. We are also expected to see further improvement if MIPUF is built using more efficient and advanced strong PUFs.
    \begin{table}[!ht]
        \centering
        \begin{tabular}{|c|c|c|c|c|}
            \hline
            Our design  & MIPUF & SHA-1     & AES       &  Overall  \\\hline
            Flip-flops         & 1,626   & 1,151   & 598    &  3,401    \\
            LUTs               & 7,028   &  1,590    & 501     & 9,219\\
            Slices              & 3,717   &  544      & 222     &  4,553 \\ 
            Block RAMs    & 0         & 0           & 3         &  3   \\
            Power($mW$) & 123.7  & 30.4     & 16.2    &  175.7   \\ \hline
        \end{tabular}
        \caption{FPGA resource and power characteristics of the hardware support of our proposed key management scheme.}
        \label{table:resource}
	\vspace*{-6mm}
    \end{table}

\section{Conclusions}
In this paper, we first proposed a novel PUF structure: MIPUF that is both secure and reconfigurable. We showcased the uniqueness, reliability, modeling attack resilience and reconfigurability of MIPUF. We then proposed a group key management scheme in IoT consists of key distribution, key storage and rekeying based on MIPUF. Security and overhead analysis on the scheme show that our design is not only secure against multiple attack methods but also low power. Our simulation result indicates that our proposed scheme spends 47.33\% less energy compared to the state-of-the-art crypto-based scheme ECPKC \cite{abdallah2015efficient} since we use low-power and energy efficient MIPUF instead of power-hungry ECC. 

\section{Acknowledgement}
This work was supported in part by the NSF under Award CNS-1513306. 

\bibliographystyle{ieeetr}
\bibliography{sigproc}

\end{document}